\documentclass[fleqn,usenatbib]{mnras}

\usepackage{mathptmx}

\usepackage[T1]{fontenc}

\DeclareRobustCommand{\VAN}[3]{#2}
\let\VANthebibliography\thebibliography
\def\thebibliography{\DeclareRobustCommand{\VAN}[3]{##3}\VANthebibliography}

\usepackage{tikz,xcolor,hyperref}

\definecolor{lime}{HTML}{A6CE39}
\DeclareRobustCommand{\orcidicon}{%
	\begin{tikzpicture}
	\draw[lime, fill=lime] (0,0) 
	circle [radius=0.16] 
	node[white] {{\fontfamily{qag}\selectfont \tiny ID}};
	\draw[white, fill=white] (-0.0625,0.095) 
	circle [radius=0.005];
	\end{tikzpicture}
	\hspace{-2mm}
}

\foreach \x in {A, ..., Z}{%
	\expandafter\xdef\csname orcid\x\endcsname{\noexpand\href{https://orcid.org/\csname orcidauthor\x\endcsname}{\noexpand\orcidicon}}
}

\usepackage{graphicx}	
\usepackage{amsmath}	
\usepackage{amssymb}	
\usepackage{color}
\usepackage{xcolor}
\usepackage{ulem}

\title[Cosmic-ray generated bubbles around their sources]{Cosmic-ray generated bubbles around their sources}

\author[B. Schroer et al.]{B. Schroer\orcidA{}$^{1,2}$\thanks{E-mail: benedikt.schroer@gssi.it}, O. Pezzi\orcidB{}$^{3,1,2}$, D. Caprioli\orcidC{}$^{4}$, C. C. Haggerty\orcidD{}$^{5}$ \& P. Blasi\orcidE{}$^{1,2}$
\\
$^1$Gran Sasso Science Institute, Viale F. Crispi 7, 67100 L'Aquila, Italy\\
$^{2}$INFN/Laboratori Nazionali del Gran Sasso, Via G. Acitelli 22, Assergi (AQ), Italy\\
$^{3}$ Istituto per la Scienza e Tecnologia dei Plasmi, Consiglio Nazionale delle Ricerche, Via Amendola 122/D, 70126 Bari, Italy\\
$^{4}$Department of Astronomy and Astrophysics, University of Chicago, 5640 S Ellis Ave, Chicago, IL 60637, USA\\
$^{5}$Institute for Astronomy, University of Hawaii, 2680 Woodlawn Drive, Honolulu, HI 96822, USA}

\date{Accepted XXX. Received YYY; in original form ZZZ}

\pubyear{2020}

\begin{document}
\label{firstpage}
\pagerange{\pageref{firstpage}--\pageref{lastpage}}
\maketitle

\begin{abstract}
Cosmic rays are thought to escape their sources streaming along the local magnetic field lines.
We show that this phenomenon generally leads to the excitation of both resonant and non-resonant streaming instabilities. The self-generated magnetic fluctuations induce particle diffusion in extended regions around the source, so that cosmic rays build up a large pressure gradient.  By means of two-dimensional (2D) and three-dimensional (3D) hybrid particle-in-cell simulations, we show that such a pressure gradient excavates a cavity around the source and leads to the formation of a cosmic-ray dominated bubble, inside which diffusivity is strongly suppressed. 
Based on the trends extracted from self-consistent simulations, we estimate that, in the absence of severe damping of the self-generated magnetic fields, the bubble should keep expanding until pressure balance with the surrounding medium is reached, corresponding to a radius of $\sim 10-50$ pc. 
The implications of the formation of these regions of low diffusivity for sources of Galactic cosmic rays are discussed. 
Special care is devoted to estimating the self-generated diffusion coefficient and the grammage that cosmic rays might accumulate in the bubbles before moving into the interstellar medium. 
Based on the results of 3D simulations, general considerations on the morphology of the $\gamma$-ray and synchrotron emission from these extended regions also are outlined.  
\end{abstract}




\section{Introduction}
\label{sec:intro}

Once accelerated, Galactic Cosmic rays (CRs) are released into the interstellar medium (ISM) in a way that is all but understood, and that depends on details of the sources. 
For instance in supernova remnants (SNRs), CRs are usually assumed to be liberated in two ways: gradually, at the instantaneous maximum energy, because of the lack of confinement in the acceleration region, and at the end of the expansion of the shell, when CRs can leave the downstream region, after substantial adiabatic losses \cite[e.g.,][]{ptuskin+05,CaprioliEscape,Cristo1}.
However the two phases, i.e. the acceleration and the escape into the ISM, are not independent.
As a matter of fact, escaping particles drive the plasma instabilities that control the confinement and the maximum energy that can be achieved by the accelerated particles \cite[]{schure,cardillo}. 
In addition, when calculating the spectra of CRs at the Earth after their transport in the Galaxy, the common assumptions are that sources release CRs in a power-law distribution and that the whole grammage accumulated by CRs in their transport throughout the Galaxy is due to propagation in the ISM. 

The propagation of CRs near their sources, where their energy density is in excess of that of the sea of Galactic CRs, is usually neglected. 
In the past few years it has become increasingly more clear that the transition between acceleration and release into the ISM requires much more attention, given its importance for different problems such as the grammage accumulated around the source \cite[]{2010PhRvDCowsik,2018APhLipari} and $\gamma$-ray emission from extended regions surrounding sources \cite[]{w28,Felix,hawc}. 

Much effort has been channelled into the investigation of this transition. In particular, it has been suggested that CRs escaping their sources along the local magnetic field can excite a resonant streaming instability \citep{Malkov,Dangelo,Nava,NavaRecchia}, which in turn confines particles close to the source due to the reduced self-generated diffusion coefficient. 
The effectiveness of this mechanism is severely limited by the damping of the Alfv\'en waves that the instability generates, as due to non-linear Landau damping \citep{nlld}, to ion-neutral damping \citep{ind} and to pre-existing turbulence \citep{Farmer_2004}. 
Due to these damping processes, the effectiveness of the confinement is likely limited to particle energies $E\lesssim$ TeV, since for higher energy particles the growth of the unstable modes is too slow. 
This limitation is qualitatively similar to the one that plagues the acceleration process: in the presence of the resonant streaming instability alone, the maximum energy at a typical SNR shock is exceedingly small \citep{Lagage1,Lagage2} and falls short of the knee by about two orders of magnitude.

A different branch of the streaming instability -the non-resonant streaming instability- \citep{Bell2004,AmatoBlasi09} has been found to grow much faster if the appropriate conditions are satisfied. 
At variance with resonant modes, the non-resonant streaming instability grows on much smaller spatial scales and is not as sensitive to the damping processes mentioned above \citep{reville+08a,zweibel+10}. 
Due to its non-resonant nature, it has the potential to reduce the diffusion coefficient by a much larger amount, provided the power in unstable modes becomes available at scales comparable with the Larmor radius of the particles in the modified magnetic field \cite[]{BlasiRev2013}. 
The effects that the development of the Bell instability in the shock precursor has on the maximum energy of the accelerated particles has been recently studied, e.g., by \cite{reville+12,caprioli+14a,caprioli+14b,caprioli+14c}.
Its role in shaping the CR spectrum released by SNRs has been investigated by \cite{diesing+21,cardillo,schure,Cristo1,Cristo2}, and, very recently, its effect in regulating the shock compression and the slope of the CRs produced via diffusive shock acceleration (DSA) has been discussed by \cite{haggerty+20,caprioli+20,diesing+21}.

The effect of the non-resonant branch on CR scattering is closely connected with the non-linear evolution of these modes: if the right conditions are satisfied (namely if there is enough energy in CRs, see below), the instability starts growing very rapidly on scales much smaller than the CR Larmor radius, so that in this stage the growing modes have little impact on particle scattering. 
As a consequence, the current of CRs exciting the instability remains only weakly affected and the growth continues, even to a stage in which $\delta B\gg B_0$. 
The power moves towards larger scales when the energy density in the growing modes becomes comparable with the energy density in the form of particles driving the CR current. 
The instability saturates when the power becomes concentrated on scales comparable with the Larmor radius of the particles, at which point the current gets disrupted, or rather confined to a smaller spatial region, closer to the location where the current originates.

While this picture has been investigated by numerous authors in the context of particle acceleration at shocks, it has recently become clear that it may also profoundly affect the transport of escaping particles in the immediate surroundings of a CR source. 
In a recent paper \citep{Schroer}, we 
have shown that the escape of high-energy (TeV) CRs from their source dramatically affects the surrounding background plasma. 
The initial propagation of these very energetic particles is quasi-ballistic, since the pathlength for parallel diffusion along the magnetic field lines is similar to the coherence scale of the Galactic magnetic field. 
This scenario naturally leads to the generation of an electric current on a spatially extended region around the source. 
If the current is strong enough, a non-resonant streaming instability is excited and the particles eventually start scattering, thereby increasing their density near the source.

Due to the diffusive motion, the initial flux tube in which CRs propagate becomes over-pressurized and starts expanding in the transverse direction. 
This eventually leads to the creation of a spherical-like bubble around CR sources, where the background plasma is partially evacuated and the diffusion coefficient is much smaller than outside the bubble. 
In principle the same chain of events can be bootstrapped by the excitation of the resonant streaming instability, but typically this process leads to smaller magnetic fluctuations, as a result of different damping processes and the typically lower growth rates of this branch.

Due to the non-linearities involved in this phenomenon, numerical simulations coupling CRs and the background plasma are crucial to properly comprehend this phenomenon. In this perspective, several works have considered a purely fluid approach in which the background plasma, treated as a magnetofluid, is dynamically affected by CRs, for which an energy balance equation is integrated \citep[e.g.,][]{2012ApJ...761..185Y, 2016A&A...585A.138D, 2017MNRAS.465.4500P, 2017ApJ...834..208R, Jiang_2018, 2019MNRAS.489..205W, 2019A&A...631A.121D,10.1093/mnras/staa2025, 10.1093/mnras/stab142}. 
This framework allows to describe larger systems, but fails to treat particle scattering self consistently; i.e., one usually has to put in by hand a value for the CR diffusion coefficient. 
Another approach is based on a combined magnetohydrodynamic (MHD) - Particle-In-Cell (PIC) approach \citep{bai+15,2018MNRAS.476.2779L,vanmarle+19}, in which the background plasma is always described within the magnetohydrodynamics (MHD) framework but CRs are an ensemble of quasi-particles self-consistently coupled to the magnetofluid. 
A third approach is kinetic \citep{2019ApJ...887..165H}, in which both the background and the CR populations are treated within the hybrid framework, i.e., protons (belonging to both the background plasma and to the CR population) are a kinetic species while electrons are a massless background fluid. 
Although neglecting the electron dynamics, this last approach properly retains the micro-physics of the problem, since electron-scale physics is usually negligible for the above phenomena. 

\citet{Schroer} have used 2D hybrid simulations to test the expectation that a bubble would be excavated in the ISM around a CR source. 
In the same paper it was shown that, for the conditions expected for a young SNR, the CRs may excite the Bell instability and indications were found of reduced diffusion coefficient. 
In this work, we extend these previous results in several ways: 
first, we describe more accurately the growth of different modes, by calculating the power spectrum of the excited fluctuations; 
second, we provide a quantitative estimate of the diffusion coefficient by propagating test particles in a snapshot of the simulation and proving that the square displacement eventually becomes linear in time. 
We also find that the diffusion coefficient for the conditions adopted here is about a few times the Bohm diffusion coefficient. 
Third, we extend the simulation to three spatial dimensions, to make sure that the results are not affected by the reduced dimensionality. 

Finally, we make an attempt to draw some conclusions about the morphology of the emission from the bubble, as due to the interactions of CRs with the gas and the magnetic fields. 
These latter results should be taken with a grain of salt in that we extrapolate our results, that refer to a relatively early phase of the bubble, to spatial scales that are appropriate for astrophysical systems. 
We also discuss in some detail the caveats that apply to our results, due to the assumptions that we are forced to make in order to make this difficult problem computationally treatable.  

The paper is organized as follows. 
In section \ref{sec:analytical} we present analytic estimates for the flux of accelerated particles escaping from a typical SNR, which suggests that the non-resonant streaming instability will occur. 
In Section \ref{sec:simulation} we describe the computational setup and in Section \ref{sec:results} we present our results of the 2D and 3D simulations and discuss their physical implications. 
Finally, in section \ref{sec:conclusion} we summarize our results and present our conclusions.
 
\section{Theoretical Background}
\label{sec:analytical}

We start by providing simple analytical estimates that set the basic background for studying the onset of CR-driven instabilities in the region surrounding a source of CRs, such as a SNR. 

The escape from the acceleration region is all but understood, and in fact the modern view of the process of particle acceleration at a non-relativistic shock is based on the fact that escape is necessary to produce the magnetic perturbations that are responsible for particle scattering close to the shock \citep{Bell2004}. 
The escape that this statement refers to is the one associated with particles that at any given time reach the maximum energy possible at that given time $E_{\rm max}(t)$. 
For an observer outside the acceleration region, the instantaneous spectrum of these particles is very peaked around the maximum energy. 
As discussed, e.g., by \cite{ptuskin+05,CaprioliEscape,Cristo1}, the total spectrum released by an individual SNR should be the sum of two contributions, that due to the time integration of the instantaneous escape and that due to all particles accelerated, advected downstream and eventually liberated at a later time, when the SNR bubble dissipates into the ISM. Below we will refer to the particles escaping from upstream at any given time, in that they provide the current density for the process we intend to describe. 

A particle of energy $E$ that escapes the acceleration region is expected to move in the turbulent magnetic field of the Galaxy. 
The only handle we have on this quantity comes from measurements of the secondary/primary ratios at the Earth, and on the measurement of the flux of unstable isotopes, such as $^{10}$Be. 
An often used estimate for the diffusion coefficient associated to such turbulent field is $D(E)\simeq 3\times 10^{28}E_{GeV}^{1/2}$ (for energies above $\sim 10$ GeV/n), although much better determinations are available in the literature \citep[e.g.,][]{CRgrammage,Evoli2020}. 
Such improvements are however inconsequential for the purpose of the qualitative argument we wish to make here. 

For the high-energy particles we are interested in, the pathlength for diffusion in the Galaxy can be estimated from $D(E)= c \lambda(E)/3$, which leads to $\lambda(E)\simeq 1~{\rm pc}~E_{GeV}^{1/2}$. 
For energies $E\gtrsim 2.5$ TeV, this becomes larger than $\sim 50\,$pc, which is the  order of magnitude of the coherence length of the Galactic magnetic field \citep{Coherence_length,Coherence_length2}. 
This means that, in first approximation,  CRs with such energies carry out a quasi-ballistic motion on scales of $\sim 50$ pc around the source, at least if the Galactic turbulent field were the only one present in that region.

Let us consider here a simple 1D model in which CRs are released as a single burst by a SNR. 
Assuming particles are injected with a power-law spectrum in momentum $p^{-4}$ ranging from $E_{\text{min}}=1\,$GeV to $E_{\text{\rm max}}=1\,$PeV, the solution for the cosmic ray flux inside the flux tube can readily be obtained as
\begin{equation}
\label{eq:flux}
    \phi_{\rm CR}(E>E_0) = \frac{L_{\rm CR}}{2\pi R_s^2\Lambda E_0},
\end{equation}
where  $R_s\sim 1-3\,$pc is the radius of the source, $E_0$ is the minimum momentum of particles in the escaping current here $\sim$ few TeV, $\Lambda=\log(E_{\text{\rm max}}/E_{\text{min}})\approx14$, and the luminosity in CRs is estimated as $L_{\rm CR}\approx \eta \frac{E_{SN}}{T_{S}}$, with $\eta\sim 10\%$ being the fraction of SN energy ending up in CRs over a timescale comparable to the Sedov-Taylor time, $T_S$. 

The condition for exciting the non-resonant branch of the streaming instability \citep[][]{Bell2004,AmatoBlasi09} then reads
\begin{equation}\label{eq:Bell}
    \phi_{\rm CR}(E>E_0) \frac{E_0}{c} \gg \frac{B_0^2}{4\pi}.
\end{equation}
For typical values of the Galactic magnetic field, $3\,\mu$G and for  $E_{SN}=10^{51}\,$erg and $T_{S}=300\,$yr, the left hand side is $\sim 54\,$eV/cm$^3$, about a factor of $\Bar{\sigma}:= 4\pi\phi_{\rm CR}(E>E_0) E_0 /(B_0^2c) = 100$ larger than twice the magnetic energy density on the right hand side, $0.2\,$eV/cm$^3$. 

Clearly this estimate should be taken as an order of magnitude calculation, since CR escape is likely more complex.
For instance, one of the assumptions involved in deriving this estimate is the spectrum of the accelerated particles at the shock, taken here as $\propto E^{-2}$. 
For the observationally-preferred \citep{CRgrammage} and  theoretically-motivated \citep{haggerty+20,caprioli+20}
steeper spectra, e.g., $E^{-2.3}$, the energy density is decreased only by a factor of $2$ at $2.5\,$TeV energies, meaning that the condition in Equation \ref{eq:Bell} would still be satisfied by a factor of 50. 

In fact, we expect such a condition to be generically fulfilled for SNRs since it depends only on the flux of escaping particles $\phi_{\rm CR}(E>E_0)$, which can be shown to be the same in the flux tube as at the SNR shock. 
Since the Bell instability is thought to be excited at SNR shocks for normal parameters it should as well be excited by the escaping particles if the magnetic field in the ISM is not that much different from the one at the shock. 
This expectation can be seen with a quick calculation:
If a fraction $\zeta$ of the ram pressure of the background gas energy density is transferred to CRs at the shock, the CR energy density reads
\begin{equation}
    \epsilon_{\rm CR} = \zeta n_{\text{gas}} m_{\text{gas}} v_s^2 = \frac{E_{\rm CR}}{\frac{4}{3}\pi R_s^3},
\end{equation}
where $n_{\text{gas}}$ is the background gas number density, $ m_{\text{gas}}$ its mass and we assumed that the energy density is the total CR energy $E_{\rm CR}$ divided by the volume of the remnant. The energy flux of CRs at the shock is then given by the velocity of the shock $v_s$ times $\epsilon_{\rm CR}$. 
A general property of DSA is that the density of particles escaping from upstream is a fraction of $\frac{v_s}{c}$ of the density at the shock, but limited to the particles with the highest momentum;
this simply follows from the assumption that CR stream freely at the speed of light after escape \cite[]{CaprioliEscape}. 
Hence, the energy flux escaping from the shock region into the tube is given by $\epsilon_{\rm CR}\frac{v_s}{c}c = \epsilon_{\rm CR}v_s$, the same as at the shock. 
From this follows directly the total energy flux of CRs in the tube using $R_s \approx v_s t$ which then gives
\begin{equation}
    \phi_{\rm CR} = \zeta n_{\text{gas}} m_{\text{gas}} v_s^2\frac{v_s}{c} c= \frac{E_{\rm CR}v_s}{\frac{4}{3}\pi R_s^3} = \frac{E_{\rm CR}}{\frac{4}{3}\pi R_s^2 t}\,.
\end{equation}
Using the normalisation of the particle spectrum $\frac{E_{\rm CR}}{t} = \int_{E_{min}}^\infty \mathrm{d}E EQ(E)$ one can obtain the flux of CRs above a certain energy which results in the same as in Equation \ref{eq:flux}, except for the geometrical factor $\frac{4}{3}$. 

The estimates above show that for rather typical conditions for the region around a SNR, the flux of escaping CRs is sufficient to excite a non-resonant streaming instability, which in turn profoundly changes the motion of the particles around the source. 

The magnetic field perturbations produced by the non-resonant instability initially grow on spatial scales that are much smaller than the Larmor radius of the particles in the current. In fact, the fastest growing mode is at a wave number  $k_{\rm max}$ given by \citep{Bell2004}:
\begin{equation}
k_{\rm max}B_{0}=\frac{4\pi}{c}J_{\rm CR}(E>E_0),
\end{equation}
with $J_{\rm CR}(E>E_0)=e \phi_{\rm CR}(E>E_0)$ being the electric current associated with particles with energy $\gtrsim E_0$. 
The growth rate of this mode is $\gamma_{\rm max}=k_{\rm max}v_{A}$ with the Alfv\'en speed $v_A = B/\sqrt{4\pi m_{\text{gas}} n_{\text{gas}}}$, which for our reference SNR parameters returns $\gamma_{\rm max}^{-1} \approx 1.1 \rm (E/2.5 TeV)\,$yr.

The magnetic field growth continues until it eventually saturates when equipartition between the energy density in CR and the magnetic field is reached or equivalently until power is transferred from small scales $k_{\rm max}^{-1}$ to the scale of the Larmor radius of particles in the amplified magnetic field \citep{Bell2004}. There is some level of ambiguity in the condition for saturation of the Bell instability. For instance, \cite{Riquelme} discussed the dependence of the saturation on the initial value of the magnetic field $B_0$. This can be an important correction when $B_0$ is very small, but for the case of SNRs in the ISM the estimate reported above is appropriate \citep[][]{zacharegkas+21p}. 

Notice that, from the estimates above, the magnetic field at saturation would only be larger than $B_0$ by the square root of the CR/magnetic energy density, i.e., a factor $\lesssim 10$. 
Nevertheless, the effects on CR transport would be prominent, in that this power would be available at scales comparable with the CR Larmor radius, while at these scales the power in the Galactic turbulence is much smaller, as can be inferred from recent estimates of the Galactic diffusion coefficient (see for instance \citep{evoli+19a,Evoli2020}). 
Hence, the diffusion coefficient characterizing the motion of the particles in the shock vicinity turns out to be suppressed with respect to the Galactic one. 
An estimate of the diffusion coefficient is $D(E)= r_{L} c/3\approx 10^{25}\rm cm^{2}/s$, i.e., several orders of magnitude smaller than expected at the same energy for Galactic CRs. 
Note that all of this is expected to happen on time scales of a few $\gamma_{\rm max}^{-1}$, much shorter than any other time scale of this problem. 
This point may be a reason of concern, in that the standard analysis of the Bell instability assumes a constant current $J_{\rm CR}$.
In our picture of the escape, CRs of a given energy escape at a given time, while at a later time particles with lower energy will contribute to the current. 

Particles of a given energy $E$ escape during a time interval which is approximately a fraction of the age of the remnant at the time they escape.
In this case, given the short value of $\gamma_{\rm max}^{-1}$ for high-energy particles, the condition is safely satisfied. 

Once the instability has reached saturation, scattering occurs with a pathlength of the order of the Larmor radius of CRs in the amplified field, which is now much smaller than the coherence scale of the Galactic magnetic field. 
As a consequence, the motion transitions from ballistic to diffusive: 
CRs are isotropized and their density increases as a result of the reduced drift velocity, the latter being proportional to the CR density gradient. 
One should recall that in the standard ISM the energy density in the form of CRs and ISM gas are basically the same. 
The effect illustrated above immediately implies that the CR energy density in the flux tube where CRs are diffusing increases way above the ISM one. 
Hence, a pressure gradient in the transverse direction arises, which acts as a force that pushes the tube sideways, thereby making the problem 2D. 
In other words, a cavity is expected to be formed with high CR density and large magnetic fluctuations inside, but with gas being swept outwards toward the edges. 

This dynamic effect will eventually dictate the overall evolution of the CRs as it influences their environment. 
It seems reasonable to assume that the expansion of the bubble driven by the CR over-pressure continues until pressure balance with the external ISM is achieved. 
This condition gives a final size of the CR bubble of $L\approx (\eta E_{s}/P_{\rm ISM})^{1/3} \sim 60\,$pc, with $P_{\rm ISM}$ the ISM pressure. Clearly, whether this condition gets to be fulfilled or not depends on several additional pieces of the problem, that are not under control. 
For instance, the argument is based upon the assumption that the magnetic field perturbations remain unabated, namely that they do not suffer damping after the action of the current has ceased. 
If this condition fails, a smaller size of the bubble should be expected. 

\section{Numerical method}
\label{sec:simulation}

To follow the non-linear dynamics of CRs in the environment around their source, we perform simulations with {\tt dHybridR}, a relativistic hybrid code with kinetic ions and massless, charge-neutralizing fluid electrons \citep{2019ApJ...887..165H}. 
{\tt dHybridR} is the relativistic extension of the Newtonian code {\tt dHybrid} \citep{2007CoPhC.176..419G} and it is suitable for properly simulating CR-driven streaming instabilities \citep{haggerty+19p,2019ICRC...36..483Z}.
The advantage of hybrid codes in comparison with fully-kinetic PIC codes is that they do not resolve small electron scales, as they are usually dynamically negligible, and therefore are better suited to self-consistently simulate the long-term coupling of CRs and background plasma.

The code solves a set of equations consisting of the Vlasov equation for different species $s$ and the Maxwell equations: 
\begin{gather}
      \label{eq:vlasov}
    \frac{\partial f_s}{\partial t} + {\bf v}\cdot {\bf \nabla} f_s+ \frac{q_s}{m_s}({\bf E} + \frac{{\bf v}}{c} \times {\bf B})\cdot {\bf \nabla}_v f_s = 0\\
    \frac{\partial {\bf B}}{\partial t} = -c {\bf \nabla} \times {\bf E\label{eq:faraday}}\\
    \frac{\partial {\bf E}}{\partial t} = c {\bf\nabla} \times {\bf B} - 4\pi {\bf J} \label{eq:ampere}\\
    {\bf \nabla} \cdot {\bf B} = 0\\
    {\bf \nabla} \cdot {\bf E} = \sum_s q_sn_s
 \end{gather}
where $f_s({\bf x}, {\bf v}, t)$ is the phase-space distribution function for a given species with charge $q_s$ and mass $m_s$, ${\bf E}$ and ${\bf B}$ are the electric and magnetic field, the total current density is given by ${\bf J} \equiv \sum_s q_s n_s{\bf V}_s$ being $n_s \equiv \int f_s d^3v$ and ${\bf V}_s \equiv \int {\bf v}f_sd^3v/n_s$ the number density and the bulk velocity of each species, respectively. In the hybrid approach electrons are a massless charge-neutralizing fluid, because their mass is negligibly small compared to the ion mass. To ensure quasi neutrality in the system the electron number density is set to $n_e=n_{\text{gas}}$ from which follows ${\bf \nabla} \cdot {\bf E} = 0$ and ${\bf J} = e n_{\text{gas}}({\bf V}_{\text{gas}}-{\bf V}_e)$.

The Vlasov equation (Eq.~\ref{eq:vlasov}) describes the evolution of ions subject to the Lorentz force. 
To solve this equation {\tt dHybridR} uses a Monte Carlo approach, i.e., the ion distribution function is approximated by a large number of macro-particles sampled from the underlying distribution function in momentum. Due to Liouville's theorem these macro-particles can then be evolved in time under the influence of the Lorentz force solving effectively 
\begin{equation}
    m \frac{\mathrm{d}\gamma {\bf v}}{\mathrm{d}t}=q{\bf E}+\frac{q}{c}{\bf v}\times {\bf B}
\end{equation}
with the relativistic Lorentz factor $\gamma=1/\sqrt{1-v^2/c^2}$ for each macro particle via the relativistic Boris algorithm. After each time step the particles' position and velocity can be interpolated onto a grid to obtain a fluid density and bulk flow for each grid cell.

To obtain the electromagnetic fields we start by using the Darwin approximation i.e. neglecting the displacement current in the Amp\`ere's law (Eq.~\ref{eq:ampere}) so that ${\bf\nabla} \times {\bf B} = \frac{4\pi}{c} {\bf J}$. 
Here we just state under which conditions this approximation is justified, for a detailed derivation see \cite{2019ApJ...887..165H}.
The approximation holds if three conditions are fulfilled. 
First, the velocity of the background population has to be non relativistic i.e. $V_{\text{gas}}\ll c$.
Second, the CR number density has to be negligible compared to $n_{\text{gas}}$ and lastly, the Alfv\'en speed must be smaller than the speed of light $v_A\ll c$.
After some algebra, the electric field is given by:
\begin{equation}
    {\bf E} = -\frac{{\bf V}_{\text{gas}}}{c}\times {\bf B}+ \frac{{\bf J}}{enc}\times {\bf B} -\frac{1}{en}{\bf \nabla}P_e \, .
\end{equation}
The only thing left to close the system of equations is to assume a relation between $P_e$ and $n_e$. While dHybridR in general allows to use different polytropic indices $\gamma_{\rm eff}$, we choose the simplest adiabatic closure so that $P_e\propto n_e^{5/3}$.

\subsection{Simulation Setup}

In simulations, physical quantities are normalized to the number density ($n_0$) and magnetic field strength ($B_0$) of the initial background plasma;
lengths are normalized to the ion inertial length $d_i = v_A/\Omega_{ci}$ (with $v_A$ the Alfv\'en speed), time to the inverse ion cyclotron frequency $\Omega_{ci}^{-1}=\frac{mc}{eB_0}$
, and velocity to the Alfv\'en speed $v_A$.
We keep all three components of the momenta and electromagnetic fields while extending the system in two or three dimensions in physical space respectively. 
The background ion temperature is set in a way that $\beta_i = 2 v_{th,i}^2/v_A^2=2$, i.e., the thermal ions gyroradius $r_{g,i}=d_i$.

For our 2D benchmark run, the simulation grid, of size $5000\times 7000\,d_i$, is discretized  in $7500\times10500$ cells (i.e., $\Delta x=\Delta y \simeq 0.66 d_i$). 
For our 3D run we use a smaller box of $1440^3$ cells with size $1200^3\,d_i^3$.
We impose an open boundary condition on the background plasma along the $x$-direction, namely the direction in which the initial CR current is, while all boundary conditions in other directions are periodic. 
CRs have open boundaries in all directions.
The thermal background plasma is described with $N_{\rm ppc}=4$ ($N_{\rm ppc}=2$ for 3D) particles per cell, a density $n_0$ and a Maxwellian distribution.  
In order to compensate for the background gas leaving the box at the boundaries in the $x$ direction, particles of the background plasma are artificially injected at these surfaces to keep the background gas density constant.
We set a background magnetic field of strength $B_0$ and directed along $x$.
CRs are injected with $N_{\rm ppc}=16$ and with an isotropic momentum distribution at the $x=0$ boundary between $y=3200\,d_i$ and $3800\,d_i$, hence only CRs with a positive $p_x$ enter the simulation box. 
They have a momentum of $p=100\,m_{\text{gas}} v_A$ and are injected with a density of $0.0133\,n_0$.
We fix $c=20\,v_A$; hence the width of the injection region corresponds to $6$ CR gyroradii and CRs have a Lorentz factor of $5$.
With these values, taking into account that only half of the CRs contribute to the current and that the density is diluted by a factor of $\sim 7$ due to the area of the current being enlarged by $2$ gyroradii because of the initial gyration, the Bell condition for the excitation of the non-resonant instability is fulfilled by a factor of $\Bar{\sigma}\simeq (n_{\text{\rm CR}}/n_0)(cv_d/v_A^2)\gamma\sim 5$. 
Note that this value is chosen rather conservatively compared to what is expected for a SNR ($\Bar{\sigma}_{\rm SNR}\sim 100$, see \S\ref{sec:analytical}).

In the 3D case the reduced box size as well as the computationally higher demanding simulation force us to use a slightly different approach. We choose to inject CRs as a cold beam along the $x$-direction with a density of $n_{\rm CR}=0.01\,n_0$ between $y=520\,d_i$ and $680\,d_i$. All the other quantities (e.g., $p$, $c/v_A$, etc...) are the same as in 2D. 
With these parameters, we have $\Bar{\sigma}_{\rm 3D}\sim 20$ (factor two from density and another factor two from drift velocity). 
This leads to an increased growth rate of the waves and allows to properly see the formation of a bubble within computationally reasonable time and length scales. 

The normalized growth rate and fastest growing wave number in simulations units can be obtained as 
\begin{equation}\gamma_{\rm max}\Omega_{ci}^{-1} = k_{\rm max}\,d_i = \frac{1}{2}\frac{n_{\rm CR}}{n_{0}}\frac{v_d}{v_A}
\end{equation}
for the non-resonant instability and $k_{\rm max}\,d_i = \frac{m v_A}{p}$ and $\gamma_{\rm max}\Omega_{ci}^{-1} = \frac{\pi}{8}\frac{n_{\rm CR}}{n_0} \frac{v_d}{v_A}$ for the resonant instability. This results in a non-resonant $\gamma_{\rm max}^{-1}\Omega_{ci} = k_{\rm max}^{-1}d_i^{-1}$ of $40$ for the 2D simulation and of $10$ for the 3D simulation. The resonant mode grows on time scales of $50\,\Omega_{ci}^{-1}$ ($12.8\,\Omega_{ci}^{-1}$) in the 2D (3D) case and on spatial scales of $100\,d_i$.

\section{Results}
\label{sec:results}

\subsection{2D Simulation}
\subsubsection{Bubble Formation and Morphology}

\begin{figure}
    \centering
	\includegraphics[width=\columnwidth]{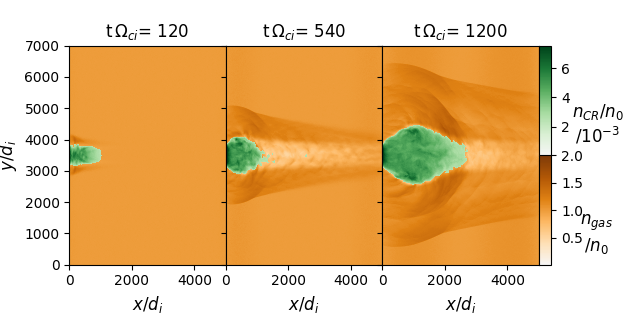}
    \caption{Background plasma density (gold) overlapped with the CR density (green, limited to the region where $n_{\rm CR} > 2.5\times 10^{-3}\,n_0$) given at three different times for the simulation corresponding to a free streaming phase (left), the formation of a bubble (center) and the expansion of the bubble (right).}
    \label{fig:snapshot}
\end{figure}

It is instructive to first look at the overall evolution of the particles before diving into the details. 
Figure \ref{fig:snapshot} illustrates the CR number density (in green) superimposed to the background plasma density (in gold). 
The CR density is only shown where it exceeds $2.5\times10^{-3}\,n_0$, solely for illustration purposes.

Let us start with noticing some properties of the particles in the injection region.
In the beginning (left panel) CRs move ballistically and gyrate around the background magnetic field (free escape), filling a flux tube of transverse size comparable to the source size.
At later times, if the propagation continued ballistically, one would expect a homogeneous cylinder whose extent keeps growing $\propto t$. 
Instead, a region of high CR density forms around the source and only few particles escape along ${\bf B}_0$, most of the CRs being isotropized by scattering off the self-generated magnetic turbulence.

The enhanced CR density produces an over-pressure in the region occupied by CRs, which acts as a force that pushes the background gas outward. 
The result is the formation of a bubble-like structure, filled with CRs and magnetic field perturbations, surrounded by an envelope of over-dense background gas (right panel of Figure \ref{fig:snapshot}). 
A similar but weaker effect can be seen due to the CRs that escape along the $x$ direction and push the background plasma out of the expected tube as the CR pressure exceeds the ambient one even for ballistic escape, as discussed in section \ref{sec:analytical}.
Despite the severe evacuation of the gas by the CRs, for our parameters  the density of CRs inside the bubble remains smaller than the gas density.

\begin{figure}
\centering
	\includegraphics[width=0.8\columnwidth]{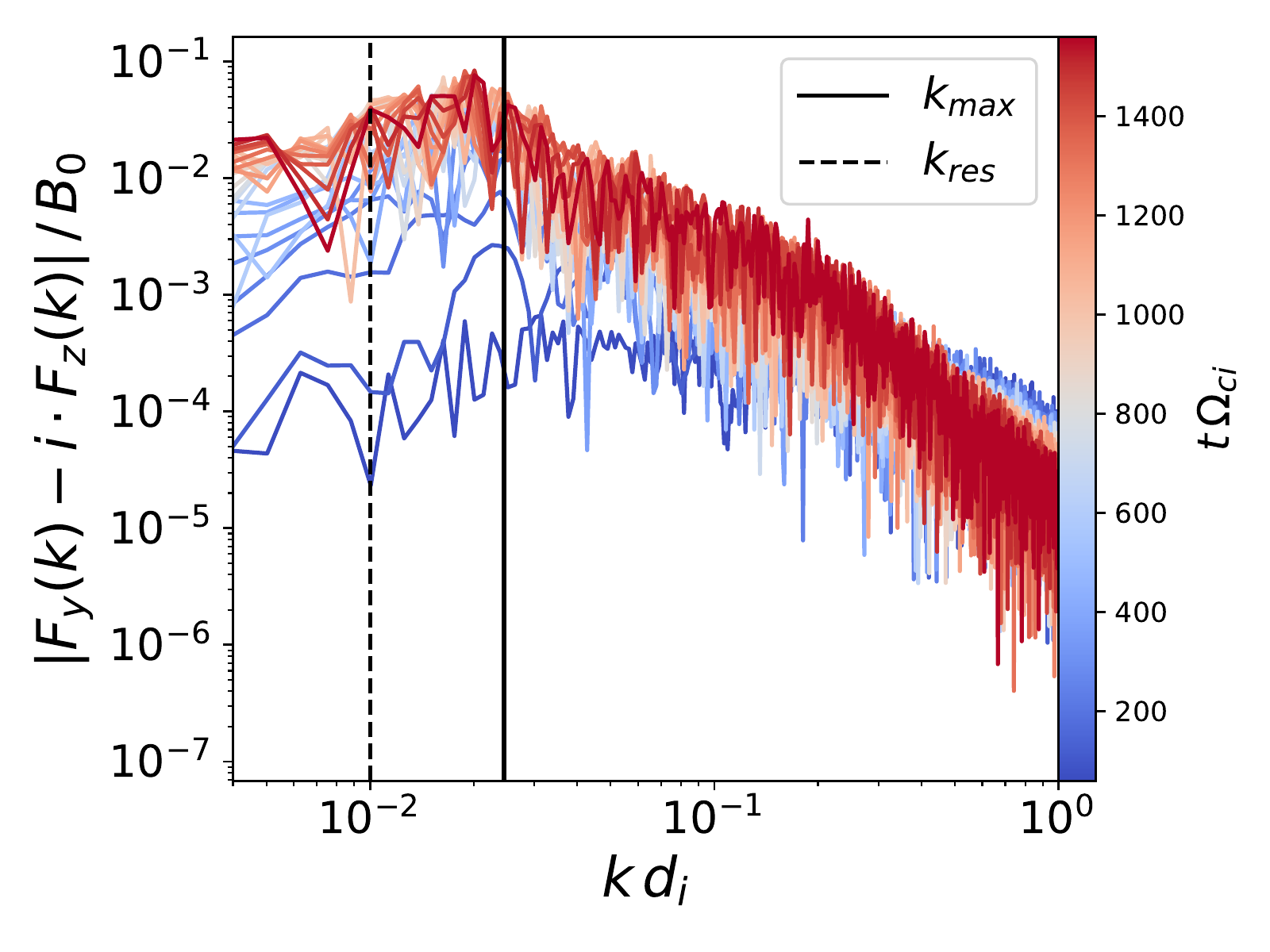}
	\includegraphics[width=0.8\columnwidth]{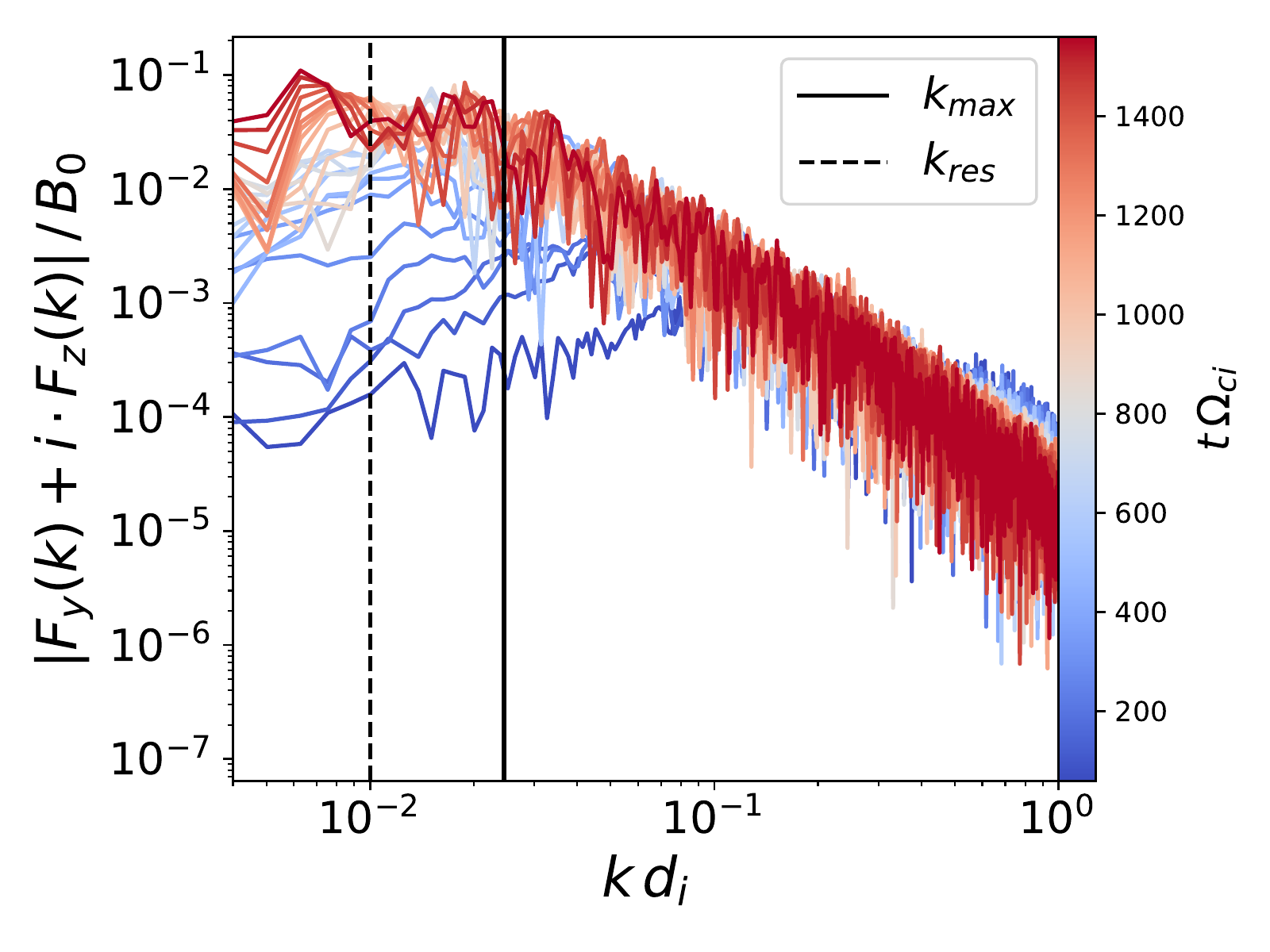}
    \caption{Fourier spectrum of the perpendicular magnetic field in left-handed (LH, top) and right-handed (RH, bottom) modes as a function of the wave number $k$. The black line shows the theoretical value of the non-resonant wave number $k_{\rm max}$ and the dotted line shows the resonant scale.}
    \label{fig:magnetic_field}
\end{figure}

\subsubsection{Nature of Self-generated Turbulence}
The particle trapping responsible for the formation of the bubble is due to CR scattering off self-generated perturbations. 
In order to quantify this phenomenon, we calculated the power spectra of left-handed (LH) and right-handed (RH) modes at different times during the simulation. 
The results are shown in Figure \ref{fig:magnetic_field}, in the form of the discrete Fourier transform $F_i(k)$ of $B_y$ and $B_z$ performed along $x$ in the combinations $B_y-\mathrm{i}B_z$ and $B_y+\mathrm{i}B_z$, respectively. 
Fields are averaged over $y$ between $y=3200$ and $y=3800$,  which corresponds to the vertical extent of the injection region.
In the same figure the vertical solid (dashed) line indicates the wave number where the fastest growing non resonant (resonant) mode is expected to develop. These expectations refer to the original current, while the actual peak of the power spectrum is expected at somewhat different values of $k$ due to the complex non-linear dynamics that develops. 

Initially, the development of the RH modes occurs at $k\sim k_{\rm max}$; then, at later times, the power starts moving toward smaller values of $k$, namely larger scales, closer to the CR gyroradius. 
During the non-linear phase the RH and LH modes lose their identity and perturbations of both types develop. In particular, the LH modes, which are the most effective in scattering CRs, have a peak close to the resonant wavenumber, $k_{\rm res}$. 

The strongest rise at $k_{\rm res}$, from $10^{-4}$ to $\approx 10^{-1}$ in magnetic power, occurs at the time of bubble formation, namely $\Omega_{ci} t\approx 420$ which corresponds to $\approx 10\gamma_{\rm max}^{-1}$, in good agreement with the estimate of the saturation time of non-resonant modes. 
At even later times the current is mainly localized in the region of the bubble, which keeps expanding, though at a lower rate. 

\subsubsection{Self-generated Diffusion Coefficient}
Inside the bubble, CRs become isotropized due to scattering off the magnetic perturbations. To quantify this phenomenon one can look at the mean pitch-angle cosine of the CR particle distribution defined as 
\begin{equation}
    \langle\mu\rangle = \frac{\int \mathrm{d}^3x \frac{p_x({\bf x})}{p} n_{\rm CR}({\bf x})}{ \int \mathrm{d}^3x n_{\rm CR}({\bf x})},
\end{equation}
where $p_x({\bf x})$ is the CR plasma momentum in the $x$ direction at position ${\bf x}$. 
Injected particles start with $\langle\mu\rangle=0.5$, as expected for a distribution that is isotropic on a half sphere. 
At $\Omega_{ci} t=1320$, the mean pitch angle cosine is reduced to $\approx 0.14$, meaning that the distribution is gradually approaching isotropy. 
The fact that it is not perfectly isotropized could be related to the fact that CRs have an open boundary condition at $x=0$, therefore particles with a negative pitch angle can leave the simulation more easily and this leakage of particles with a large negative pitch angle leads to a non-zero positive mean $\mu$. Alternatively it could mean that not all the particles are diffusing which to an extent is certainly true for the particles in the initial flux tube outside the bubble.
To a certain extent both of these effects play a role in determining $\langle\mu\rangle$. Nonetheless, it remains true that there is a net trend toward isotropy, which shows that the majority of CRs are isotropized and effectively trapped near the source.
The remaining current of particles is marginally stable w.r.t. the instability, i.e., it has a $\Bar{\sigma} \sim 1$ and can therefore survive. 

To further quantify the diffusion of particles inside the bubble we take a snapshot of the magnetic field at $\Omega_{ci} t=1320$ in our simulation and perform a test particle simulation in such a field configuration, using a simple Boris particle pusher to evolve the position of $20,000$ particles. 
The diffusion coefficient is estimated as $D_{xx}(t)=\langle\Delta x^2\rangle/(2t)$, $\langle\Delta x^2\rangle$ being the mean displacement of the particles along one direction \citep[see, e.g.,][]{caprioli+14c}.
For our test particle simulation we see that, after an initial phase of free streaming, $D_{xx}(t)$ becomes time independent and converges to a $\sim 6000\,d_i^2\Omega_{ci}$, corresponding to a few times the Bohm diffusion coefficient. 

An independent upper limit on the diffusion coefficient can be obtained by imposing that the particles remain in the bubble at a given time $t$.
This means that the diffusion coefficient should be smaller than the square of the bubbles size in the $x$ direction divided by the time $t$. 
This estimate returns a comparable estimate of $\sim 5300\,d_i^2\Omega_{ci}$.
Clearly, both estimates are to be taken with caution. A possible caveat that applies to the first method is that the simulation of the transport of test particles was carried out in the stationary field of the snapshot, hence it neglects the time dependence of the background turbulence 
As for the second method, as we stressed above, it has to be considered as an upper limit rather than a quantitative estimate. Yet, the fact that the two estimates are comparable and that there is evidence of particle isotropization seem to support the general picture outlined above. 

\subsubsection{Bubble Expansion and Saturation}
The onset of the diffusive phase inside the bubble coincides with the increase in the pressure of CRs and the establishment of a pressure gradient with respect to the outside region, which in turn drive the expansion of the bubble. 
In principle, the expansion may continue until pressure equilibration is achieved. Unfortunately it is not possible to follow the dynamics of the bubbles evolution for such a long time in our simulations because density waves launched in the background plasma reach the $y$-boundaries and re-enter the box before balance is achieved. 

\begin{figure}
\centering
	\includegraphics[width=\columnwidth]{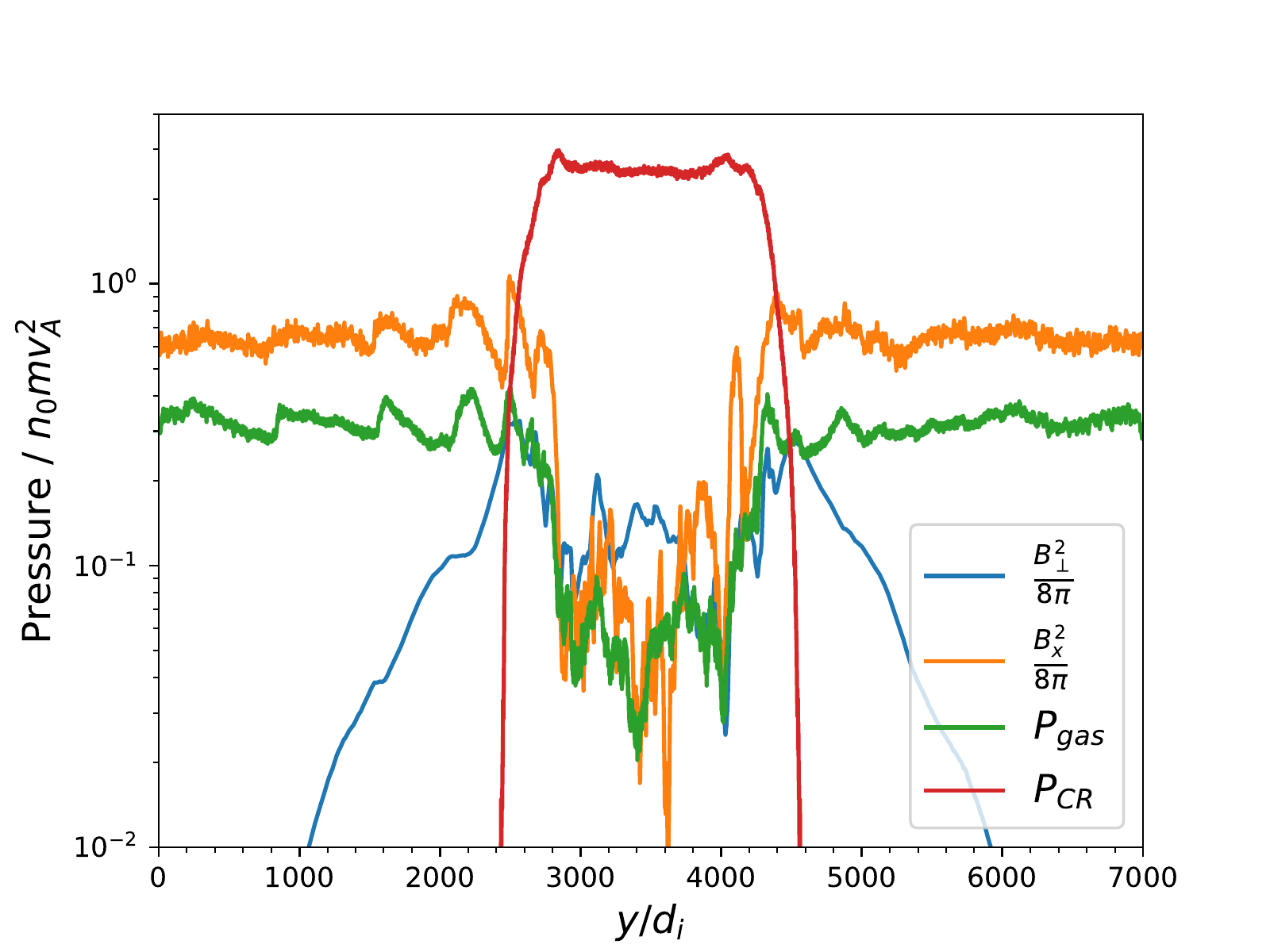}
    \caption{Pressure in CRs, background plasma and magnetic field as a function of $y$ at time $t\Omega_{ci}=1500$ averaged from $x=1200$ to $x=2100$.}
    \label{fig:pressure}
\end{figure}

In Figure \ref{fig:pressure} we show the pressure of the three different components, CRs, background plasma and magnetic field. The latter is split in the pressure of the background field $B_x$ and that of the mostly turbulent component $B_\perp$. These pressures are computed in a cross section of the simulation box along $y$ at a late time $\Omega_{ci} t=1500$ averaged between $x=1200$ and $x=2100$.

The bubble can be identified as the region in which the CR pressure exceeds all of the other pressure components. This figure demonstrates that the bubble has very well defined boundaries in $y$ as any CR particle trying to leave the bubble will eventually gyrate back into it due to swept up tangential magnetic field lines draping the bubble which can be seen in the magnetic pressure components.
Furthermore, it is evident that the over-pressure in CRs evacuates the bubble of its gas content leading to low gas pressure inside the bubble. This situation may have potentially important phenomenological implications, as CRs in the bubble may interact with gas and accumulate grammage. We discuss this point in more detail in \S\ref{sec:gamma}.

The overpressurized CR bubble is also responsible for a partial evacuation of the magnetic field, at least the ordered field along $x$. 
Part of this field is compressed on the edges and pushed in the perpendicular direction, as can be seen in the blue curve in Figure \ref{fig:pressure}. 
Most of the transverse field is however in the form of turbulent field with a typical magnitude of $\sim 0.3\,B_0$. 
In fact the turbulent part inside the bubble seems to have all three components of the magnetic field, as expected in the case of strong turbulence that leads to quasi-Bohm diffusion.

In passing, we point out that recent X-ray observations  of the extended region around Geminga \citep{Geminga_xray},  spatially coincident with its TeV halo, hint at the fact that in that region the magnetic field is appreciably lower than in the ISM, perhaps suggesting that we are observing a bubble similar to the one discussed above. 
In fact, the case of pulsar wind nebulae (PWNe) is expected to be different in that the total current should be negligible. However, it is also possible that processes of charge separation \citep{buccia} and/or a net charge \citep{gupta+21} may create regions where the non-resonant instability may be excited. 
In addition, the formation of the bubble and its expansion are in fact a generic product of the diffusive motion of particles. 
In principle these conditions may be produced also when other instabilities are excited, although the growth can be slower.

Finally, one may notice that the pressure of the external medium, set at unity in the beginning of the simulation, is reduced in Figure \ref{fig:pressure}. This is an artifact of the open boundary condition at $x=0$, which allows for gas particle leakage from the box. We partially address this problem by using an injector that forces particle number density conservation in the box. The scheme however does not keep the pressure constant, hence with time the gas component gets cooler. This imperfect treatment of the gas behavior should not affect the conclusions concerning the formation of the overpressurized CR bubble illustrated above. 

Another possible shortcoming of the simulation is that the open boundary condition for CRs on the left-hand side, i.e., close to the source, allows CRs to leave the box on that side. 
In reality, CRs would go back towards the source and might even return from that region. This introduces effectively a leakage of CR pressure to the left which is not expected in a more physical picture of the system. 
In order to investigate the effect of this leakage we tested the other extreme of setting the boundary as reflective, i.e., every particle trying to leave the system on the left gets reflected back into the bubble eliminating the leakage completely.
Within this setup an increase in CR pressure is observed, as expected, but the overall evolution of the quantities stays the same. While there is a stronger over-pressure and a slightly larger expansion of the bubble, we did not observe any new feature. 
Hence, we conclude that the leakage of CR pressure to the left does not influence the conclusions drawn above.

\subsection{3D Simulation}

\begin{figure*}
\centering
	\includegraphics[width=0.33\textwidth]{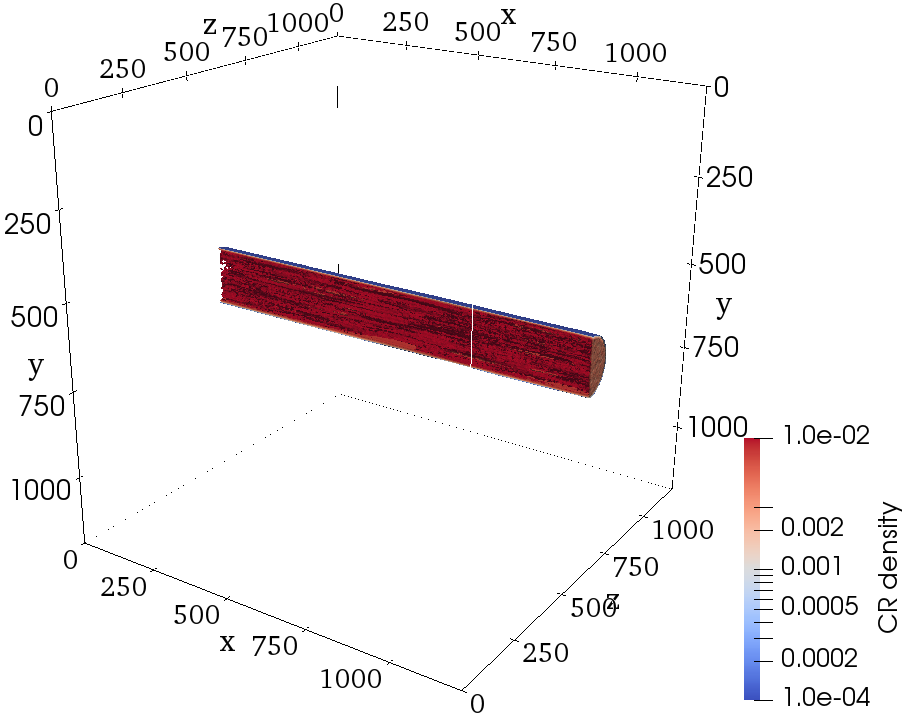}
    \includegraphics[width=0.33\textwidth]{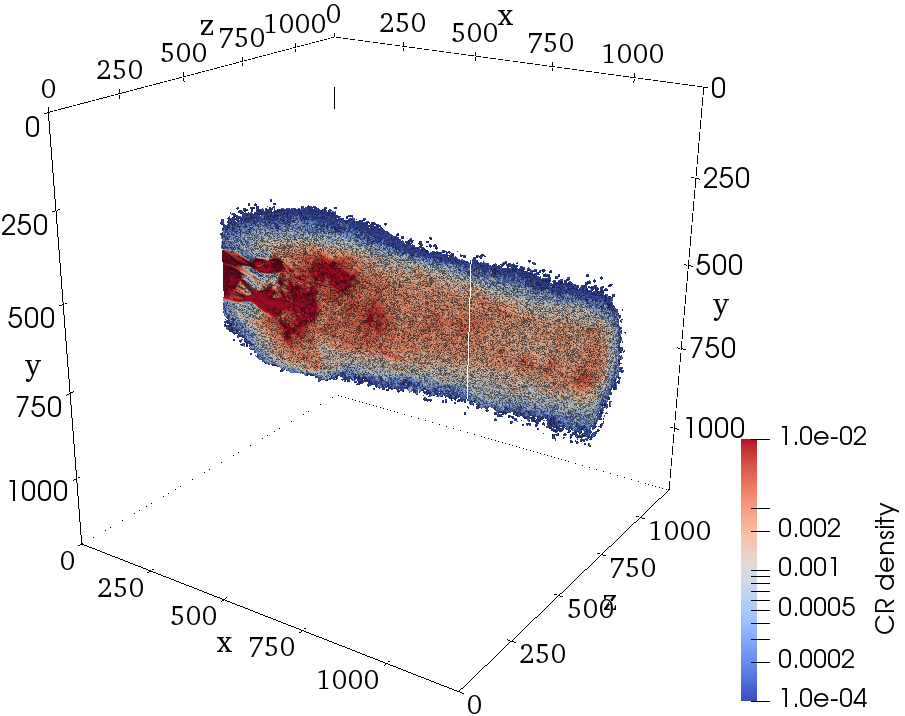}
	\includegraphics[width=0.33\textwidth]{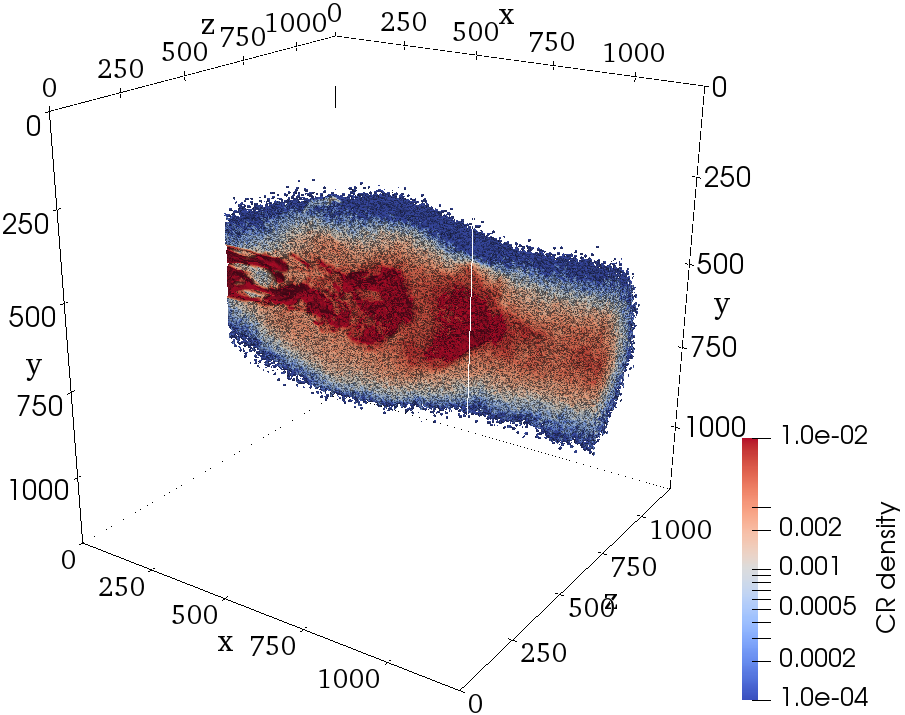}
    \caption{3D Contour plots of CR density at $t\Omega_{ci}$ equal to $60$, $180$ and $300$ for the 3D simulation.}
    \label{fig:3D_contour}
\end{figure*}

The results discussed above, and partially outlined in our previous article \cite[]{Schroer}, leave room to the possibility that at least some of them may be due to the reduced dimensionality of the 2D simulations used so far. 
In order to address this possibility we ran 3D simulations of the same problem using {\tt dHybridR}. 
Due to the much higher computational requirements of these simulations, we reduced the box size to $1200\times1200\times1200\,d_i$ divided in $1440\times1440\times1440$ cells, and tracked the beam of CRs only until shortly after the formation of the bubble-like structure when the density waves in the outside plasma reach the periodic boundary in $y$- and $z$-direction.

Figure \ref{fig:3D_contour} shows 3D contour plots of the CR density at three different times, the viewing angle is chosen in order to correspond to Figure \ref{fig:snapshot} for the 2D case with the injection of CRs on the left hand side.

\begin{figure}
\centering
	\includegraphics[width=\columnwidth]{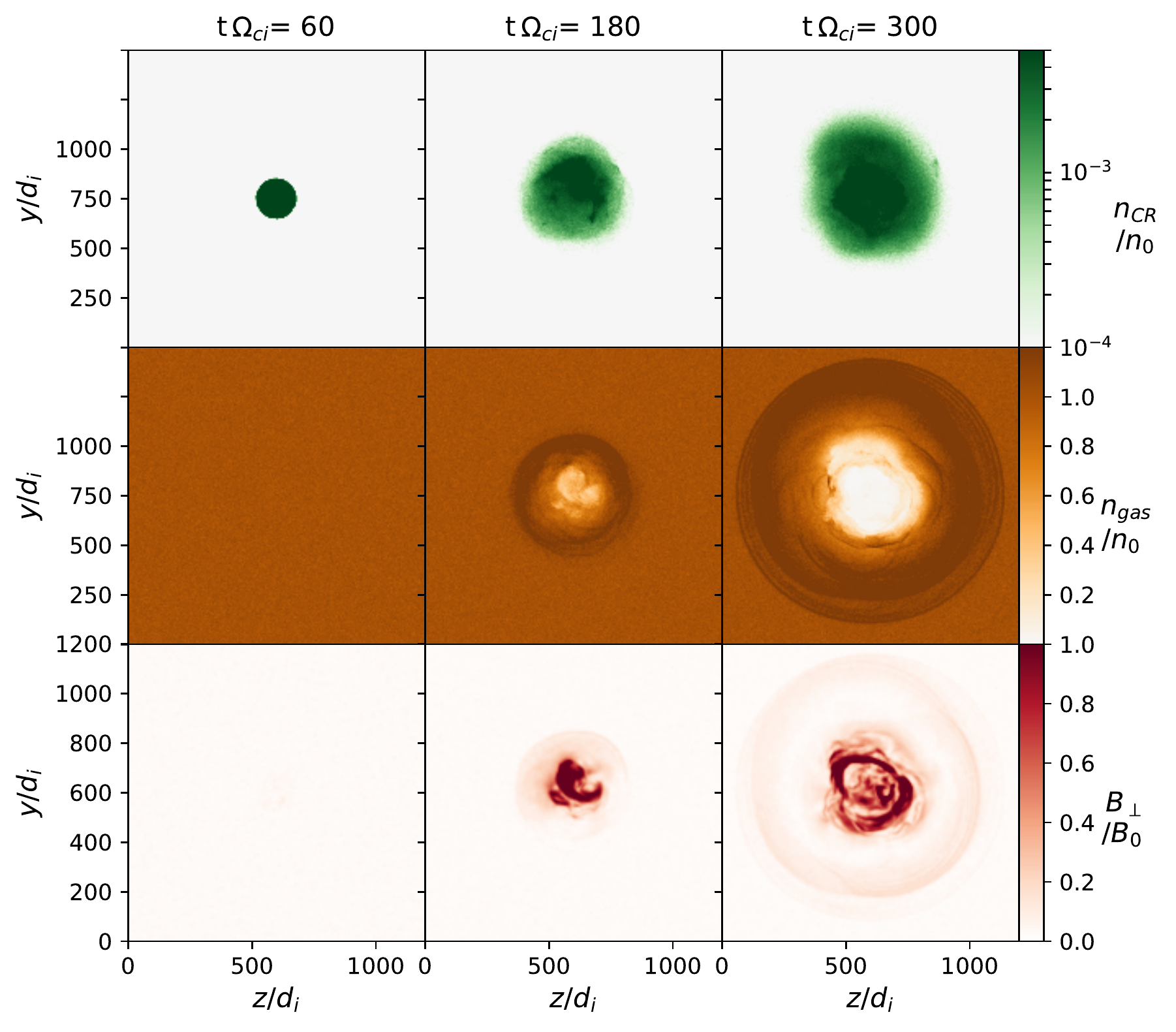}
    \caption{CR density at two different times and plasma density in the $y-z$-plane for the 3D simulation. All quantities are averaged along $x$ between $500\,d_i$ and $520\,d_i$ to represent a cross section through the bubble.}
    \label{fig:3D}
\end{figure}

Though the box is smaller, the evolution looks very similar to the 2D case, in that CRs start scattering off magnetic self-produced fluctuations and form an expanding flux tube. 
Since in 3D we inject CRs as a cold beam, there is no initial gyration, but once particles start scattering off the magnetic fluctuations, the beam opens up by one gyroradius to each side; 
this can be seen comparing the left with the middle panel of Figure \ref{fig:3D_contour} and Figure \ref{fig:3D}. Furthermore, it is evident that particles get trapped close to the injection region, this can be seen by the red region in the middle panel which is located close to the injection region.
At later times the expansion of the tube is purely due to the over-pressure in CRs, as can be inferred by comparing the middle and right panel of Figure \ref{fig:3D_contour}.
The over pressurised red region expands in all directions, although it expands preferentially along the $x$ direction which is mainly due to the CRs being injected as a cold beam along $x$, therefore pushing the bubble from the left hand side. Interestingly, also in 3D particles start to scatter effectively at roughly $10\,\gamma_{\rm max}^{-1}\approx 100\Omega_{ci}^{-1}$, as in the 2D case.

In 3D we can check the transverse evolution of the bubble excavated by CRs, namely its expansion in the $y-z-$ plane. 
Figure \ref{fig:3D} displays the CR density, the background gas density and the perpendicular magnetic field at early, intermediate, and late times averaged along the $x$ axis between $500\,d_i$ and $520\,d_i$ taken from Figure \ref{fig:3D_contour}, well inside the bubble.

Figure \ref{fig:3D} shows that the transverse expansion of the CR-filled bubble is quasi-spherical, as in 2D simulations. 
Compared with the 2D case, the evacuation of central region of the bubble is somewhat stronger, most likely as a consequence of the larger initial CR over-pressure. 
We notice that the interface between the bubble and the external medium is less sharp in 3D than in 2D, highlighting a more pronounced mixing between the two media. 
This mixing also causes the presence of gas structures inside the bubble itself, as shown, e.g., in the top right part of the excavated bubble in the central-right panel of Figure \ref{fig:3D}. 
Similar to the 2D case, there is a strong turbulent magnetic field inside the bubble which moves with the expanding bubble but seems localized in a slightly smaller region.

Additionally, in 3D there are two ring-like structures visible in the magnetic field. The outer one is the result of the compression of the magnetic field at the boundary between external medium and compressed gas expelled from the cavity. 
The inner ring is harder to associate with structures, but might reflect an inward moving wave due to the interaction with the external medium (similar to a reverse shock).

\begin{figure}
    \centering
	\includegraphics[width=0.8\columnwidth]{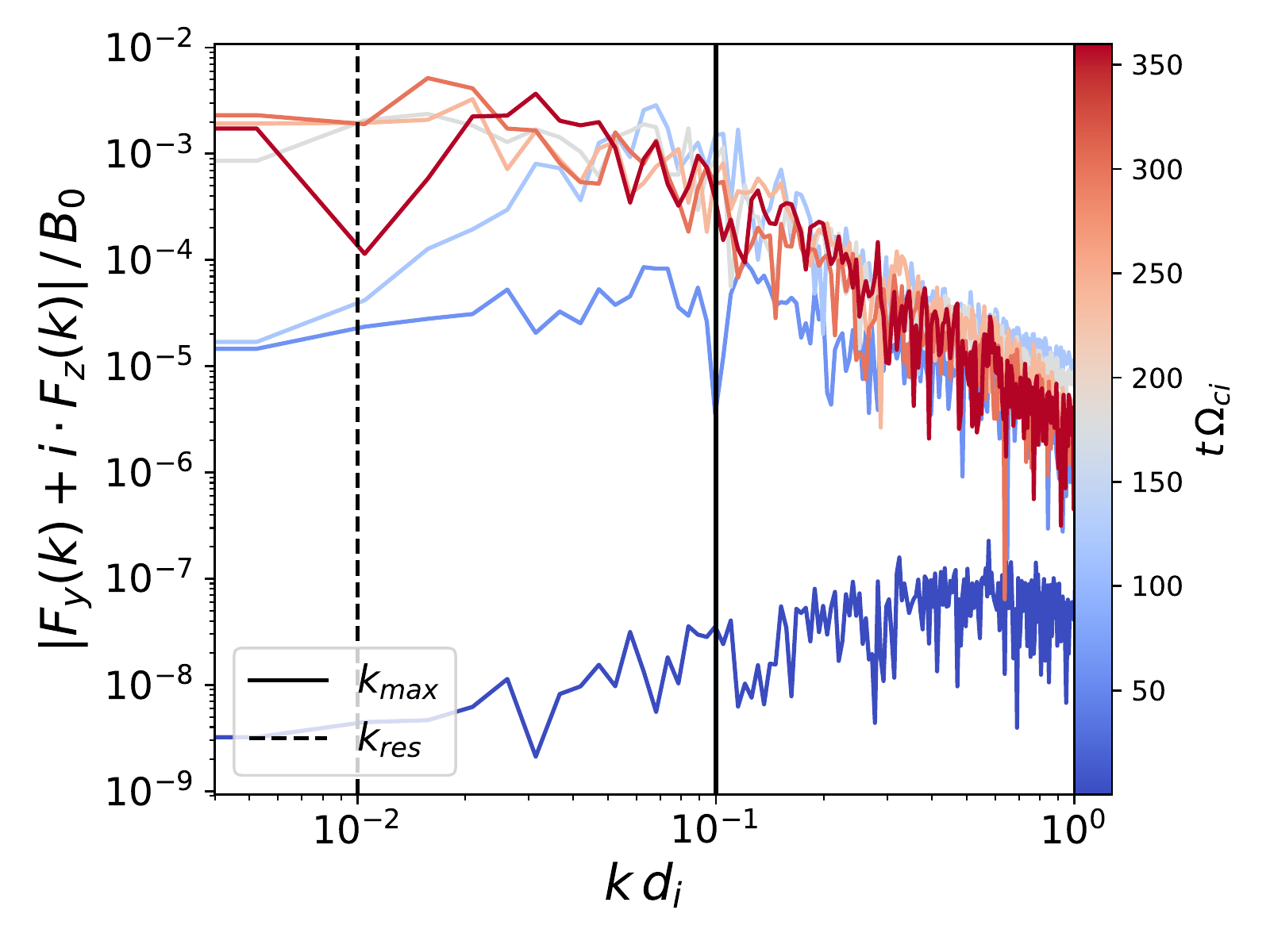}
	\includegraphics[width=0.8\columnwidth]{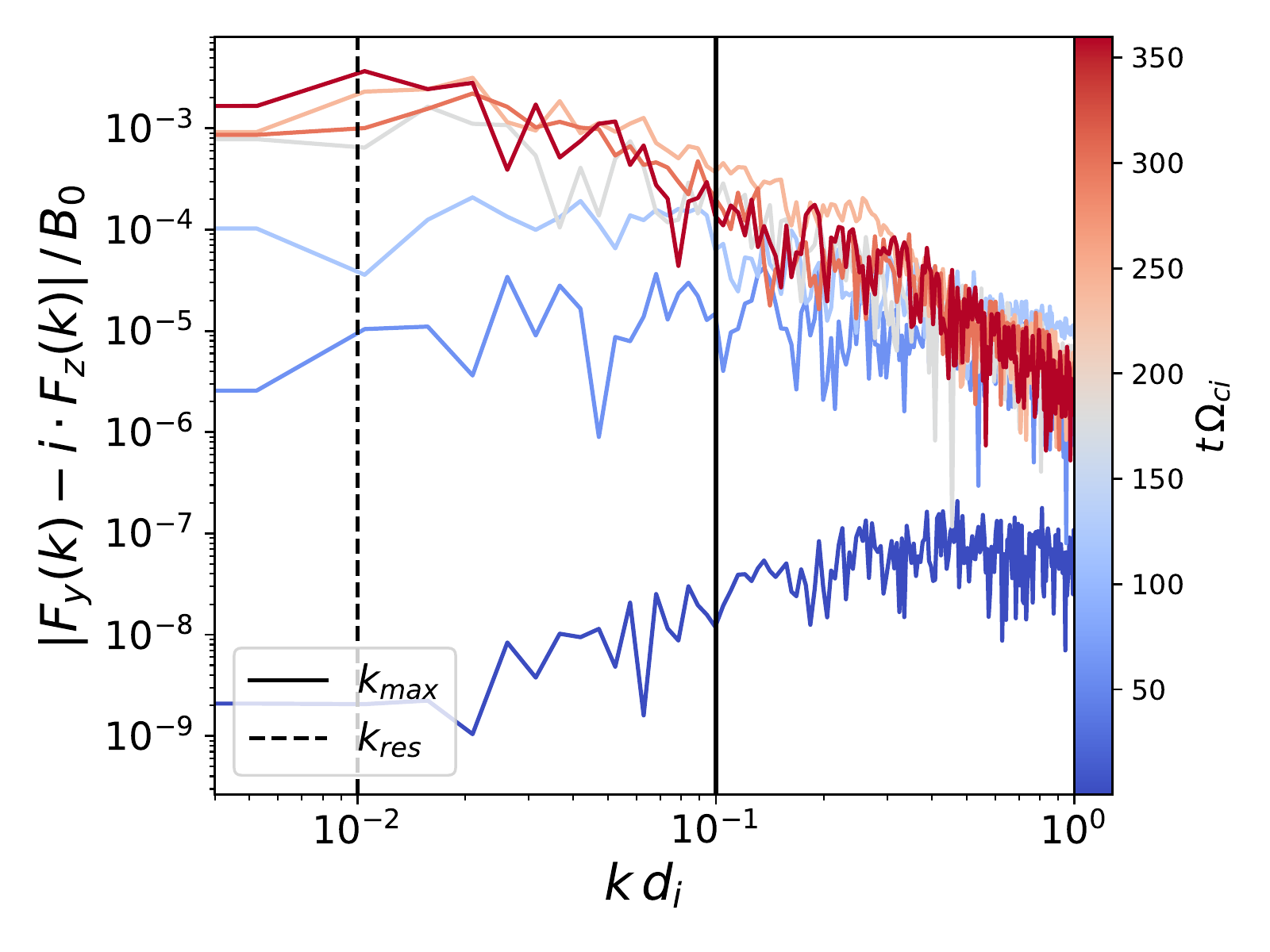}
    \caption{Fourier spectrum along $x$ of the right-handed (top panel) and left-handed (bottom panel) magnetic field modes inside the whole 3D simulation box as a function of the wave number $k$. The black line shows the theoretical value of the fastest growing wave number $k_{\rm max}$ of the non-resonant mode and the dashed line corresponds to the resonant wave number.}
    \label{fig:3D_fourier}
\end{figure}

Apart from these small differences the 3D simulation reproduces the physical picture emerging from the 2D simulation. 
The structure of the self-generated magnetic field can be appreciated in Figure \ref{fig:3D_fourier}, which shows the magnetic power in RH (top) and LH (bottom) modes as a function of wave number $k$ at different times for the 3D simulation. 
In 3D the magnetic fields are first integrated over the $y$ and $z$ directions in the whole simulation box, and then Fourier transformed along $x$. 
Due to the initialization as a cold beam the maximum growing wave number is closer to the predicted one. 
After a time $\sim10\,\gamma_{\rm max}^{-1}\approx 100\,\Omega_{ci}^{-1}$ one can see the saturation and cascading to the larger scales, accompanied by an increase in power at the resonant wave number in both LH and RH modes, which enables strong particle scattering and results in the formation of the expanding bubble of CRs. 

This suggests that the main physical aspects of the problem can be captured by 2D  simulations and, at the same time, puts the conclusions drawn from the 2D simulation on firmer grounds. 
Nonetheless, small differences like the larger overlap of the CR bubble and the background gas and small structures of gas inside the bubble provide important hints for the production of hadronic $\gamma$-rays that can be expected in this scenario.

\subsection{CR interactions in the bubble}
\label{sec:gamma}

As mentioned in \S\ref{sec:intro}, in addition to the general question of understanding the plasma physics of CR self-confinement around their sources, there is observational interest in this problem, because of the recent evidence for regions of extended $\gamma$-ray emission, e.g., from around the Geminga and Monogem pulsars \cite[]{hawc} and the W28 SNR \cite[]{Gabici}. 

Both cases suggest that the diffusion coefficient in the region surrounding these sources is suppressed by about two order of magnitude with respect to the Galactic one and are likely not isolated.

The cases of PWNe and SNRs are different from the point of view of the physical processes taking place around the source.
 In the case of SNRs, we know that there must be a net current of CRs (predominantly protons) leaving the source \citep{Cristo2}. In this case all the considerations listed above apply and our investigation confirms that there should be a region of reduced diffusivity and enhanced CR density around the sources. 
In the case of PWNe, using naive standard assumptions a net current is not to be expected since these sources mainly produce equal numbers of positrons and electrons.
On the other hand, \cite{buccia} proposed that at least the highest energy pairs may escape from different locations. If so, then in some regions around a PWN a net current may be present, although it remains to be seen whether the current is large enough to excite a non-resonant streaming instability. 
The idea of a net current is further supported by PIC simulations of aligned rotators which show an excess of high-energy positrons leaving the system \citep{cerutti+15}.
As for the resonant streaming instability, it appears to be too slowly growing to explain observations \cite[]{2018PhRvD..98f3017E}. 

At present it is not known whether the small diffusion coefficient around PWNe is due to the pairs themselves or rather to the turbulence produced by the CR protons associated with the parent SNR, like the bubbles described in this paper.
The main evidence of suppressed diffusion around a SNR comes from the $\gamma$-ray emission from molecular clouds at different distances from W28 \cite[]{w28,Gabici}. 
A somewhat less clear case is that of SNR G106.3 + 2.7 \cite[]{Yiwei}, where also a small diffusion coefficient was invoked. 

\begin{figure}
\centering	\includegraphics[width=\columnwidth]{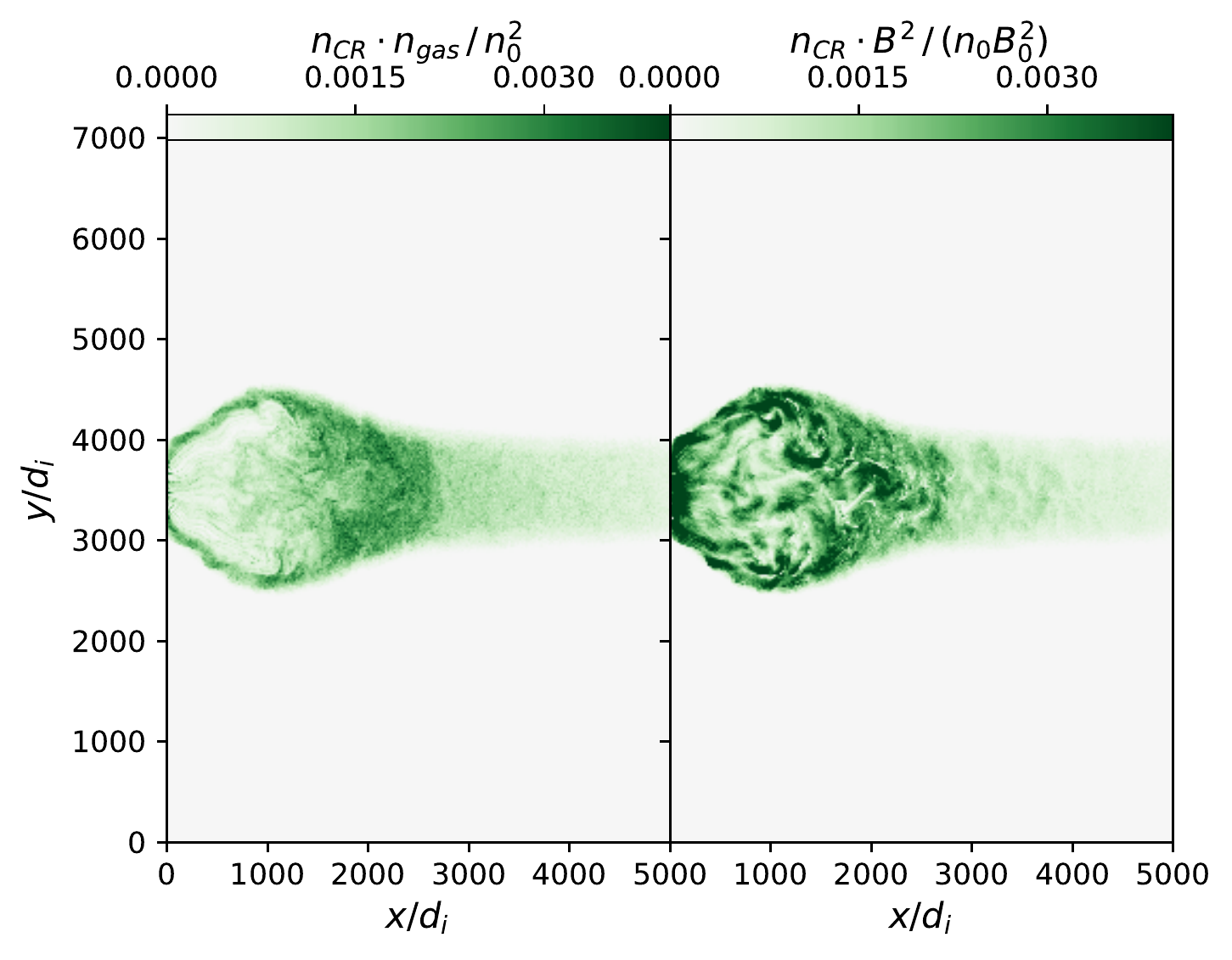}
	\includegraphics[width=\columnwidth]{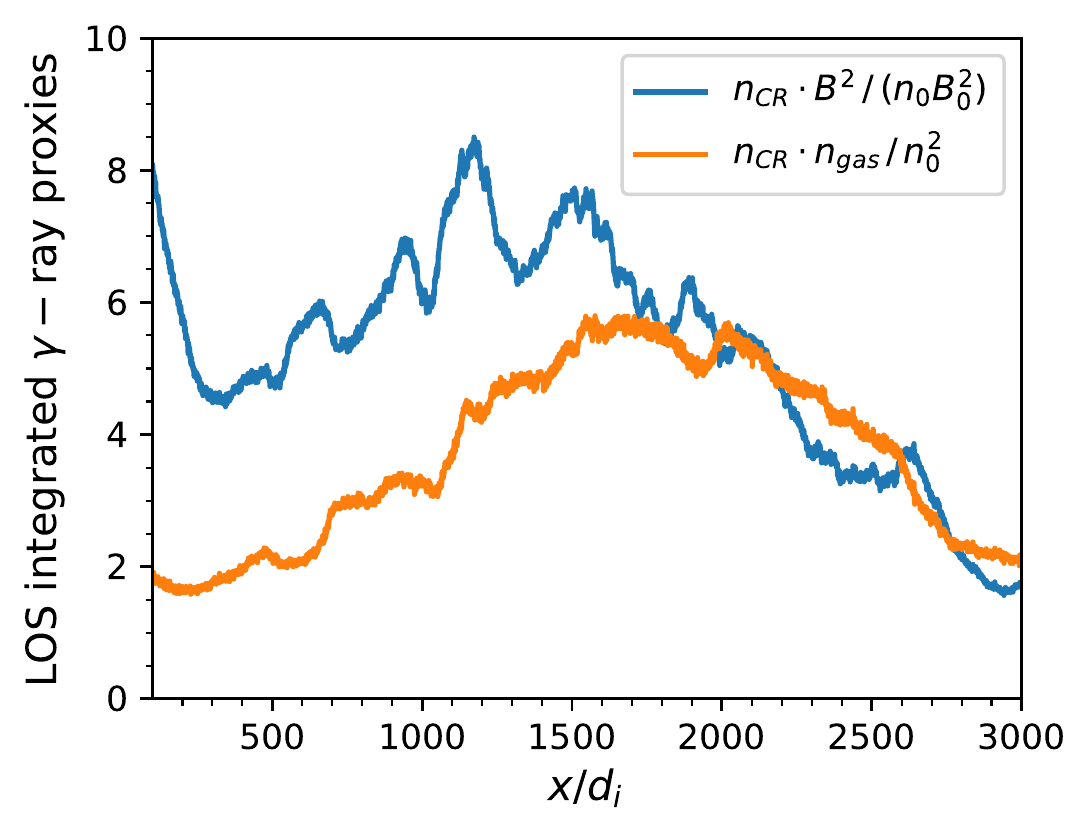}
    \caption{Product of cosmic ray density $n_{\rm CR}$ with background gas density $n_{gas}$ (top left panel) and total magnetic field (top right panel) to illustrate the probable morphology of expected $\gamma$-ray emission of the bubble via pion production or synchrotron emission. On the bottom we see this morphology integrated along the line of sight along $y$.}
    \label{fig:gamma}
\end{figure}

Although most of the circum-SNR $\gamma$-rays are expected to come from occasional molecular clouds, it may be instructive to investigate the morphology of the hadronic $\gamma$-rays that should be produced by self-generated bubbles.
We consider the quantity $n_{\rm CR}n_{gas}$, integrated along the line of sight, as a proxy of such emission. 
This quantity is plotted in the top left panel of Figure \ref{fig:gamma} for a source observed at 90 degrees orientation w.r.t. the flux tube orientation. Here we used the results of our 2D simulations, which allow us to consider a larger size of the bubble.   
One can see that, since the bubble is filled with CRs but depleted of thermal plasma, most of the $\gamma$-ray emission comes from its edges.
A molecular cloud sitting inside the bubble, instead, would probe the region where the CR density is rather uniform and the diffusion coefficient is suppressed.

Leptons liberated by a source are expected to go through the same self-generated turbulence, possibly produced by the current of CR hadrons. 
For completeness, Figure \ref{fig:gamma} (top right panel) shows the integral over lines of sight of the product $n_{\rm CR}B^2$, which is a proxy for the synchrotron emission from the same region. 
Since the bubble is almost uniformly filled with CRs and self-generated magnetic fields, its emission appears to have a much more regular morphology, less concentrated at the edges of the bubble than the $\gamma$-ray emission. 
These findings are illustrated in the bottom panel of Figure \ref{fig:gamma}, where the emission has also been averaged over the $y$ direction. 
It is important to remember that we do not expect CR escape to amplify the background magnetic field by orders of magnitude, contrary to what happens at the shock;
therefore, the contrast of synchrotron emission between the bubble and the surrounding ISM should be ascribed mostly to the (moderate) CR overdensity. 
Radio and X-ray emissions may help shed light on this phenomenon.

The existence of regions of suppressed diffusivity around SNRs is potentially  important not only for $\gamma$-ray observations, but even for quantifying the grammage that CRs may accumulate before eventually being injected into the ISM. 
The standard model of CR transport in the Galaxy allows us to infer halo size, diffusion coefficient on several kpc scales and hence diffuse $\gamma$-ray emission, CR density and other quantities, only to the extent that the grammage measured through secondary/primary ratios (for instance B/C) and secondary/secondary ratios (such as Be/B) is accumulated throughout the Galaxy. 
On the other hand, the problem is considerably more complex if some fraction of such a grammage is accumulated inside or around the sources.  

The implications range from profound revisitations of the standard model \cite[]{Cowsik,2010PhRvDCowsik,2018APhLipari}, when a large fraction of the grammage is due to near-source transport, to mild effects on the B/C and antiproton fluxes \cite[]{2019MNRAS.488.2068B}, when the grammage accumulated near the source is small. 

To estimate the importance of the near-source grammage, let $D_{\rm gal}(E)$ be the diffusion coefficient on Galactic scale, and $D(E)=\xi D_{\rm gal}(E)$ the near-source diffusion coefficient, suppressed by an amount $\xi\lesssim 1$ with respect to the Galactic one. 
We parametrize the density near the source as $n_{bkg}=\eta n_d$, where $n_d$ is the density of the Galactic disc. 
The near-source grammage equals the Galactic one if 
\begin{equation}
    \frac{\xi}{\eta}\lesssim \frac{L^2}{H h}\approx 3\times 10^{-3} \left(\frac{L}{50 pc} \right)^2 \left(\frac{H}{5 kpc} \right)^{-1}\left(\frac{h}{150 pc} \right)^{-1},
\end{equation}
where $H$ and $h$ are respectively the halo and the disc thickness. 
For $\eta\sim 1$, namely in the absence of an evacuation of the near-source region, and for $\xi\sim 10^{-2}$, one would infer that about $10\%$ of the grammage would be contributed by the near-source regions, an effect that would be probed with high-precision data from AMS-02. 
In this simple estimate we assumed that the energy dependence of the near-source and the Galactic diffusion coefficient is the same, which is all but guaranteed. Currently we have no reliable information on how the near-source diffusion coefficient depends on energy. 
If $\eta<1$, as our calculations suggest, as a result of the partial evacuation of gas from the CR blown bubble, then the result above becomes less constraining, in the sense that the near-source grammage decreases. 
On the other hand, if the diffusion coefficient in the region around the source is Bohm-like, then one should expect that at low energies the grammage accumulated by CRs before escaping into the ISM may be not negligible, although the phenomena discussed above are mainly extended to a large region around the source only for particles that can move ballistically in the beginning, and this is the case only for $E\gtrsim$ TeV. 

We finally comment on the fact that although the self-confinement of CRs in the near-source region may be a reason of concern for the grammage accumulated by CRs, the escape time from the CR blown bubbles remains much shorter than the escape time from the Galaxy. 
Hence, the standard calculations of the abundance of unstable isotopes, such as $^{10}\rm{Be}$, and the corresponding estimates of the propagation time in the Galaxy are expected to remain valid. 
On the other hand, if the scattering properties of CR electrons are similar to those of protons in the near-source region, the suppressed diffusivity in the same region may have important implications for the spectrum of leptons released by powerful CR sources. This will be discussed in a forthcoming publication. 

\section{Conclusions}
\label{sec:conclusion}

We put on solid theoretical grounds the prediction that the escape of high-energy CRs from a SNR can strongly affect the CR scattering properties in the region around the source, through the excitation of the non-resonant streaming instability. 
The phenomenon is triggered by CRs with high enough energies that their path-length in the Galactic diffusion coefficient is comparable with the coherence scale of the Galactic magnetic field, say $\sim 10-100$ pc. 
However, it has been shown \citep{Dangelo,Nava,NavaRecchia} that even resonant scattering can be enhanced due to CRs streaming around a SNR, at least for particles with $E\lesssim 1$ TeV.

The phenomenon described above should be rather general and implies that around any Galactic CR source that is powerful enough there should be an extended region where the magnetic field turbulence is enhanced, and the CR scattering consequently suppressed. 
CR escape and self-confinement around sources are intertwined mechanisms necessary to build a theory of acceleration in CR sources.

When self-generated waves have grown enough, scattering transforms the particle motion from quasi-ballistic to diffusive, thereby leading to an increase in the CR density near the source. 
At this point the CR pressure in what is usually pictured as a flux tube of cross section comparable to the source radius becomes much larger than in the surrounding ISM, hence the tube expands sideways. 

In fact, both \cite{Schroer} and the present article focus on the case in which a flux tube can be clearly identified to start with, namely a situation in which the coherence scale of the turbulent field in the Galaxy is larger than the size of the source. In particular when the source is located inside the disc region, a larger level of turbulence may be expected, and possibly a smaller value of $L_c$. One can picture this situation as one in which different parts of the shock where particles get accelerated have a quasi-parallel or a quasi-perpendicular configuration. In the regions where the field lines are approximately parallel to the shock normal the picture of particle escape should be similar to the one illustrated above, although the details of this scenario require further assessment, in terms of diffusion coefficient and final size of the expanding bubble.

The main physical picture discussed above was already proposed in our previous article \cite[]{Schroer}, while here we generalize and formalize it in several ways:

1) We extended the hybrid simulations to three spatial dimensions, to make sure that our main physical results are not affected by the dimensionality of the problem. The main difference is the morphology of the gas distribution in the bubble: 
in the 3D case there is more mixing between the plasma inside and outside the bubble, but this does not change the global 2D picture. 
This is to be expected, since turbulent mixing and fluid-like instabilities, e.g. Rayleigh-Taylor and/or Kelvin-Helmholtz, are expected to be more efficient in a 3D configuration. The mean gas density in the bubble remains appreciably smaller that the initial one, confirming that gas is pushed out by the CR overpressure. 

2) We calculated the power spectrum of the excited modes, both in 2D and 3D, to show that they initially grow on scales much smaller than the Larmor radius of the particles in the current, consistent with the expectations for Bell non-resonant modes. 
On time scales of the order of $\sim 10\gamma_{\rm max}^{-1}$ the magnetic power moves toward larger spatial scales and eventually reach the Larmor radius, at which point scattering becomes effective and particles get trapped in the near-source region. 
It is also worth noticing that in the non-linear stage both polarizations of the growing modes are present, as expected \citep[][]{haggerty+19p}. 

3) We measured the effect of the enhanced CR scattering in three independent ways. First, we calculated the mean value of the particles' pitch angle, which shows a clear signature of particle isotropization.
Second, we measured the diffusion coefficient by performing simulations of the transport of test particles in a snapshot of the hybrid run, obtaining a value few times larger than Bohm. 
Third, we considered the extent of the bubble as a function of time, also obtaining a consistent value.  

4) We elaborated on the expectation of the expansion of the CR blown bubble at later times, pointing out the caveats that apply to the description of such phase using hybrid numerical simulations. 
For the typical energy input of a SNR, it is expected that pressure balance corresponds to a final bubble size of $\sim 50$ pc. 
These estimates should be taken with caution in that they rely on an extrapolation of the results of our simulations to scales (spatial and temporal) that are much larger than what we can afford to cover at the present time or in the foreseeable future, at least with hybrid simulations. 

5) We discussed the implications of the formation of the CR blown bubble for extended $\gamma$-ray and radio observations of the circum-source region and for the CR grammage accumulated in the near-source region.
While the self-confinement of CRs near SNRs seems consistent with the evidence of  reduced diffusivity around W28 \cite[]{w28,Gabici}, it is less straightforward to apply it to the cases of PWNe such as Geminga and Monogem \cite[]{hawc}. 
The main limitation in doing so is related to the fact that PWNe are expected to release mainly electron-positron pairs, with approximately null electric current. Hence, to zero order, it is expected that the non-resonant instability cannot be excited.  
On the other hand, at the highest energies electrons and positrons may leave the PWN from different regions \cite[]{buccia,e-p-escape}, and an asymmetry between positron and electron acceleration has been reported in PIC simulations of pulsar magnetospheres \citep[e.g.,][]{cerutti+15}.
Lepton-driven instabilities \citep[e.g.,][]{bret+10}, while possibly different from proton-driven ones, can still excite non-resonant modes \cite{gupta+21} and in general produce turbulence that reduces the local diffusion coefficient with respect to typical ISM values. 

Some general implications of the scenario discussed in the present article are that the CR blown bubble should be characterized by a mostly turbulent magnetic field, while the regular filed is expected to be swept out toward the edge of the bubble. 
Inside the bubble the gas density is also expected to be lower than in the undisturbed ISM. Interestingly, a recent analysis of the X-ray emission around Geminga suggests that the magnetic field in the same region where $\gamma$-ray emission comes from is lower than average \citep{Geminga_xray}.

Probably the most important implication of the existence of regions of reduced diffusivity around sources is the possibility that while propagating around the source CRs may accumulated grammage, which eventually reflects in the production of secondary nuclei, antiprotons and positrons. 
Depending on the magnitude of the effect, this may lead to a rewriting of the standard picture of CR transport \cite[]{Cowsik,2018APhLipari} or to some sizeable effects that may appear at high energies in the B/C or in $\bar p/p$ ratios. Although the possibility that these cocoons exist has been proposed several times, mainly to address some anomalous findings such as the rising positron fraction, a theoretical motivation has been lacking. 
Recent work \cite[]{Nava,Dangelo,NavaRecchia,Schroer} and the present article are steps in that direction.

\section*{Acknowledgements}
Simulations were performed on computational resources provided by the University of Chicago Research Computing Center, the NASA High-End Computing Program through the NASA Advanced Supercomputing Division at Ames Research Center, and XSEDE TACC (TG-AST180008). DC was partially supported by NASA (grants NNX17AG30G, 80NSSC18K1218, and 80NSSC18K1726) and by NSF (grants AST-1714658, AST-1909778, and PHY-2010240) and CH by NSF (FDSS grant AGS-1936393). PB was partially funded through grant ASI/INAF n. 2017-14-H.O.
We are grateful to the anonymous referee for several interesting comments that helped us improve the quality of the manuscript.

\section*{Data Availability}
The data underlying this article  will be shared on reasonable request to the corresponding author.



\input{output.bbl}



\bsp	
\label{lastpage}
\end{document}